\documentclass[prd,aps,showpacs,tightenlines,nofootinbib,preprint,preprintnumbers]{revtex4}
\usepackage{graphicx}
\usepackage{amsfonts}
\usepackage{amssymb}
\usepackage{latexsym}
\usepackage{color}
\input{colordvi.tex}

\newcommand{\CO}{{\cal O}}

\newcommand{\bear}{\begin{array}}  \newcommand{\eear}{\end{array}}
\newcommand{\bea}{\begin{eqnarray}}  \newcommand{\eea}{\end{eqnarray}}
\newcommand{\beq}{\begin{equation}}  \newcommand{\eeq}{\end{equation}}
\newcommand{\bef}{\begin{figure}}  \newcommand{\eef}{\end{figure}}
\newcommand{\bec}{\begin{center}}  \newcommand{\eec}{\end{center}}
\newcommand{\non}{\nonumber}  
\newcommand{\lmk}{\left(}  \newcommand{\rmk}{\right)}
\newcommand{\lkk}{\left[}  \newcommand{\rkk}{\right]}
\newcommand{\lhk}{\left \{ }  \newcommand{\rhk}{\right \} }

\newcommand{\del}{\partial}  

\newcommand{\bib}{\bibitem} 
\newcommand{\la}{\left\langle} \newcommand{\ra}{\right\rangle}

\newcommand{\gsim}{\mathop{}_{\textstyle \sim}^{\textstyle >}}

\def\IB#1#2#3{{\bf #1}, #2 (19#3)}
\def\IBB#1#2#3{{\bf #1}, #2 (20#3)}
\def\IBID#1#2#3{{\it ibid}. {\bf #1}, #2 (19#3)}
\def\IBIDD#1#2#3{{\it ibid}. {\bf #1}, #2 (20#3)}

\def\APJSS#1#2#3{Astrophys. J. Suppl. Ser. {\bf #1}, #2 (20#3)}

\def\CQG#1#2#3{Class. Quantum Grav. {\bf #1}, #2 (19#3)}
\def\CQGG#1#2#3{Class. Quantum Grav. {\bf #1}, #2 (20#3)}

\def\JCAPP#1#2#3{J. Cosmol. Astropart. Phys. #1 (20#3) #2}

\def\JHEP#1#2#3{J. High Energy Phys. #1 (19#3) #2}
\def\JHEPP#1#2#3{J. High Energy Phys. #1 (20#3) #2}

\def\NPB#1#2#3{Nucl. Phys. {\bf B#1}, #2 (19#3)}

\def\PLB#1#2#3{Phys. Lett. B {\bf #1}, #2 (19#3)}
\def\PLBB#1#2#3{Phys. Lett. B {\bf #1}, #2 (20#3)}
\def\PLBold#1#2#3{Phys. Lett. {\bf#1B}, #2 (19#3)}

\def\PRD#1#2#3{Phys. Rev. D {\bf #1}, #2 (19#3)}
\def\PRDD#1#2#3{Phys. Rev. D {\bf #1}, #2 (20#3)}

\def\PRL#1#2#3{Phys. Rev. Lett. {\bf#1}, #2 (19#3)}
\def\PRLL#1#2#3{Phys. Rev. Lett. {\bf#1}, #2 (20#3)}

\def\PRT#1#2#3{Phys. Rep. {\bf#1}, #2 (19#3)}

\def\PTP#1#2#3{Prog. Theor. Phys. {\bf #1}, #2 (19#3)}

\def\PTPS#1#2#3{Prog. Theor. Phys. Suppl. {\bf #1}, #2 (19#3)}

\begin{document}
\preprint{FTPI-MINN-08-05,UMN-TH-2536/08,UT-08-02}
\title{New D-term chaotic inflation in supergravity and leptogenesis}

\author{Kenji Kadota}
\affiliation{William I. Fine Theoretical Physics
Institute, University of Minnesota, Minneapolis, MN 55455}

\author{Teruhiko Kawano}
\affiliation{Department of Physics, University of Tokyo, Hongo, Tokyo 113-0033,
Japan}

\author{Masahide Yamaguchi} 
\affiliation{Department of Physics and Mathematics, Aoyama Gakuin
University, Sagamihara 229-8558, Japan}

\date{\today}

\begin{abstract}
We present a new model of D-term dominated chaotic inflation in
supergravity. The F-flat direction present in this model is lifted by the 
dominant D-term, which leads to chaotic inflation and
subsequent reheating. No cosmic string is formed after inflation because
the U(1) gauge symmetry is broken during inflation. The leptogenesis scenario
via the inflaton decay in our D-term chaotic inflation scenario is also
discussed.
\end{abstract}


\maketitle

\section{Introduction}

\label{sec:int}

Among many types of inflation models proposed so far, chaotic inflation
is special in that it can take place around the Planck time and make
the universe large enough to avoid the recollapse \cite{chaotic}, unless
the universe is open at the beginning. On the other hand, other
inflation models occur typically at much later times so that they suffer
from the flatness (longevity) problem \cite{inflation}, namely, why the
universe lives so long starting around the Planck scale till the low energy 
scale without the collapse. 
Furthermore, other types of inflation except
chaotic and topological inflation also suffer from the initial condition
problem \cite{inflation,hi}, that is, why the inflaton field is
homogeneous over the horizon scale and takes a value which leads to a
successful inflation.

For the analysis of a chaotic inflation model, one would need to
consider the supergravity which would govern the dynamics of the early
universe \cite{SUSY}. It, however, has been long considered challenging
to realize chaotic inflation in supergravity simply because the F-term
potential of a scalar field in supergravity has an exponential growth
which prevents an inflaton from having an initial value much larger than
the reduced Planck scale $M_{p} \simeq 2.4 \times 10^{18}$~GeV.  Chaotic
inflation is still possible in supergravity, however, by allowing a
non-minimal K\"ahler potential \cite{GL,MSYY}, even though it would be
hard in general to justify a specific form of K\"ahler potential unless
some symmetry such as the Nambu-Goldstone-like shift symmetry is
introduced as done by Kawasaki, Yanagida and one of the present authors
(M.Y.) for the natural chaotic inflation in supergravity \cite{KYY}
\footnote{The application of such a shift symmetry enables us to realize
not only chaotic inflation \cite{KYY,KYY2} but also its variants in
supergravity \cite{variSG,variSG2} and superstring \cite{variSS}.}.

Another possibility to circumvent the above difficulty stemming from the
F-term is to use the D-term dominated inflaton potential because a
D-term does not have an aforementioned dangerous exponential factor.  In
the conventional models of D-term inflation \cite{dtermorig,dterm}, the
energy density is sourced by the constant Fayet-Iliopoulos (FI) term
$\xi$ in D-term while the inflaton trajectory follows the F-flat
direction whose slope is induced by the one-loop corrections.  Since the
one-loop corrections cannot exceed the tree level potential energy
density, the whole potential energy density is at most $\xi^2 \ll M_p^4$
so that the inflation cannot start from the Planckian energy scale in
such conventional D-term inflation models dominated by the FI term. They
thus necessarily give rise to hybrid inflation.  On the other hand, in
the model of D-term chaotic inflation proposed by two of the present
authors (K.K. \& Y.M) \cite{KY}, the gauge non-singlet inflaton field
identified with the (almost) F-flat direction is automatically lifted by
the dominant D-term. Such a direction can thus naturally lead to D-term
dominated chaotic inflation. It however cannot lead to the successful
reheating in the original model because the gauge symmetry under which
the inflaton is charged is unbroken and the charge conservation
prohibits the coupling between the inflaton and the standard model
fields. In this paper, we present a simpler model of the D-term chaotic
inflation in supergravity which can induce the successful reheating
where the gauge symmetry is spontaneously broken by the non-vanishing
vacuum expectation value (VEV) of the inflaton. As a concrete
illustration of a new D-term chaotic inflation scenario, we discuss the
leptogenesis \cite{FY} by the inflaton decay into the right-handed
Majorana neutrinos \cite{LGinfdec,KYY2} accompanied by the sufficient
reheating.

In the next section, we present a new model of chaotic inflation in
supergravity and discuss its dynamics and primordial fluctuations. In
Sec. \ref{sec:lepto}, we discuss the leptogenesis via the inflaton
decay. The final section is devoted to the summary and discussion.

\section{New D-term chaotic inflation model in supergravity}

\label{sec:chao}

We introduce three superfields $S, X, \overline{X}$ charged under $U(1)$
gauge symmetry and (global) $U(1)_R$ symmetry. The charges
of the superfields are 
listed in Table \ref{tab:charges}, 
which ensure our 
model is anomaly free \cite{KK,try}. 
\begin{table}[btc]
  \begin{center}
    \begin{tabular}{| c | c | c | c | c | c | c |}
        \hline 
                   & $S$ & $X$ & $\overline{X}$ & $N_i$ & $H_u$ & $L_i$ \\
        \hline
        $U(1)$        & 0   & $+1$& $-1$           & 0     & 0     & 0 \\
        \hline 
        $U(1)_R$      & $+2$& 0   & 0              & $+1$  & $+1$  & 0 \\
        \hline
    \end{tabular}
    \caption{The $U(1) \times U(1)_R$ charge assignments for the superfields.}  
    \label{tab:charges}
  \end{center}
\end{table}

The general renormalizable superpotential for these fields is then given
by
\beq
  W = \lambda S (X \overline{X} - \mu^2),
  \label{eq:super}
\eeq
where we can set the constants $\lambda$ and $\mu$ to be real and positive
for simplicity. Note that non-renormalizable terms $S(X\overline{X})^n$
can appear in the superpotential. However, as shown later,
$|X\overline{X}| \sim \mu^2$ during inflation so that such higher terms
are negligible as long as $\mu \ll M_p$.

This leads to the following scalar potential consisting
of the F-term $V_F$ and D-term $V_D$, 
along with the canonical K\"ahler potential $K(\Phi_i,\Phi_i^{\ast}) =
\sum_{i} |\Phi_i|^2$ and the minimal gauge kinetic function 
$f_{ab}(\Phi_i) = \delta_{ab}$, 
\bea
  V &=& V_F + V_D, \non \\ 
  V_F  &=&
       \lambda^2 e^{K} \lkk \,
         \biggl| X\overline{X} - \mu^2 \biggr|^2
         (1 - |S|^2 + |S|^4) \right. \non \\
       && \left.
          + |S|^2 \lhk
          \biggl| \overline{X} + X^{\ast}(X\overline{X} - \mu^2)  \biggr|^2
          + \left| X + \overline{X}^{\ast}(X\overline{X} - \mu^2)  \right|^2
               \rhk \,
            \rkk
        , \non \\
  V_D&=& \frac{g^2}{2} 
       \lmk |X|^2 - |\overline{X}|^2 \rmk^2,
\eea
where $g$ is the coupling constant 
of the U(1) gauge interaction 
and we do not introduce 
the FI term for simplicity. Here and hereafter we set the reduced
Planck scale $M_p$ to be unity and use the same symbols for the superfields
and corresponding scalar fields unless stated otherwise.

The minima 
of the F-term (the F-flat condition) 
are given by
\beq
        X\overline{X} - \mu^2 = 0, 
\qquad 
              S = 0,
\eeq
and the minima 
of the D-term (the D-flat condition) are 
given by
\beq
    |X| = |\overline{X}|.
\eeq
The global minima 
of the potential hence 
are given by
\beq
  S = 0,\,\,\, X = \mu e^{i\theta},\,\,\, \overline{X} = \mu e^{-i\theta},
\eeq
where the phase $\theta$ 
can be set to zero by the U(1) gauge transformation. 

One should notice that this superpotential and the corresponding scalar
potential are the same as those of the conventional F-term hybrid inflation
\cite{Fhybrid} in which the gauge singlet field $S$ plays the role of
an inflaton while $X$ and $\overline{X}$ remain zero 
during the inflation and then roll down to the global 
minima after the inflation. 
In order for the hybrid inflation to start,  
the field $S$ has to be relatively large but smaller than the reduced Planck 
scale $M_p$ due to the exponential factor in the F-term while $X$ and 
$\overline{X}$ almost vanish \cite{hybinit}.
 
The notable difference between the new D-term chaotic inflation and 
the conventional F-term inflation is the initial condition. 
In the new model, inflation occurs when 
$|X|\,\gsim\,1$ or $|\overline{X}|\,\gsim\,1$ 
with $S \sim 0$ and $X\overline{X} \sim \mu^2$ which almost satisfy 
the F-flat condition. 
When the universe starts
around the Planck scale, the potential energy as well as the kinetic
energy is expected to 
be of order the Planck energy density. This requires the almost
F-flat condition because  all the fields quickly roll down to
the global minimum because of the exponential factor, as will be seen below. 
The almost F-flat direction is thus naturally realized around the Planck scale. 
The potential is consequently
dominated by the D-term potential, and thus chaotic inflation 
takes place.

We now investigate 
the dynamics of the new D-term chaotic inflation model in details.  
Despite the $e^K$ factor of F-terms, due to the presence of 
the relatively small but
non-vanishing D-terms, the actual inflaton trajectory is slightly
deviated from the exact F-flat direction and given by solving the
equations (1) $\del V / \del S^{\ast} = \del V / \del
\overline{X}^{\ast} = 0$ or (2) $\del V / \del S^{\ast} = \del V / \del
X^{\ast} = 0$, depending on the initial conditions ((1) for $|X| \gg 1$
or (2) for $|\overline{X}| \gg 1$). Note that the system is invariant
under the interchange of $X$ and $\overline{X}$ so that both solutions
(trajectories) lead to the essentially same dynamics. We therefore 
concentrate on the first trajectory given by
$\overline{X} = \overline{X}(X), S=0$, and we call this trajectory T.

We here confirm that inflation indeed occurs along this (almost F-flat) 
field trajectory.  For this purpose, we first evaluate the mass terms of
the field $S$ along the trajectory T, $V_{,ij}|_{T}
\phi_{i}^{\ast} \phi_{j}$ with $V_{,ij} \equiv \del^2
V/(\del\phi_i^{\ast} \del\phi_j)$. Here $\phi_{i}$ represents $S$, $X$,
or $\overline{X}$. The suffix $T$ represents the evaluation along the
trajectory $T$. Then, the mass matrix of the fields $S$, $V_{,ij}|_{T}$, is
given by
\bea
  V_{,SS}|_{T} &\simeq&
      \lambda^2 e^{K} (|X|^2 + |\overline{X}|^2), \non \\
  V_{,SX}|_{T} &=& 0, \quad  V_{,S\overline{X}}|_{T} = 0.
  \label{eq:matrixS}
\eea
The effective squared mass of the field $S$ is much larger than the Hubble
parameter squared $H^2 \simeq g^2 |X|^4/2$ unless the constant $\lambda$
is exponentially small, which makes $S$ quickly go to the zero. As a
result, we can safely set $S$ to be zero and we can discuss the dynamics
of the inflaton based on the following potential,
\beq
  V_{\rm eff}(X,\overline{X}) 
    \equiv V(X,\overline{X},S=0) 
    = \lambda^2 e^{K} |X\overline{X} - \mu^2|^2 
                     + \frac{g^2}{2} (|X|^2 - |\overline{X}|^2)^2.
\eeq
By use of the $U(1)$ gauge symmetry, we can, for instance, make the
field $X$ real without loss of generality, so that the ${\rm Im}
\overline{X}$ rapidly goes to the zero because the effective mass
squared of the imaginary part of $\overline{X}$ is given by $m^2_{{\rm
Im} \overline{X}} \simeq \lambda^2 e^{K} X^2$. 
We therefore consider the
following effective potential, by redefining the fields $X \equiv
\sqrt{2}\, {\rm Re}\,X$, $\overline{X} \equiv \sqrt{2}\, {\rm
Re}\,\overline{X}$ (we take both $X$ and $\overline{X}$ to be
positive for definiteness) and $\mu' \equiv \sqrt{2} \mu$,
\beq 
  V_{\rm eff}(X,\overline{X}) = 
     \frac{\lambda^2}{4} e^{K} 
        \lmk X\overline{X} - \mu'^2 \rmk^2 
      + \frac{g^2}{8} \lmk X^2 - \overline{X}^2 \rmk^2
\eeq
with $K = (X^2 + \overline{X}^2) / 2$ and the canonical kinetic terms.

The dynamics of the inflation can be discussed based on the above
potential. As mentioned before, we consider the case that $X \gg 1$
initially, which implies $\overline{X} \ll 1$ for $\mu' \ll 1$. For such
a case, the trajectory (T) is characterized by the condition $\del V /
\del \overline{X} = 0$ which is equivalent to
\bea
  X\overline{X} - \mu'^2 &=& 
    e^{-K} 
    \frac{g^2 \lmk X^2-\overline{X}^2 \rmk \overline{X}}
         {\frac{\lambda^2}{2} X + \frac{\lambda^2}{4} \overline{X} 
          \lmk X\overline{X} - \mu'^2 \rmk}
              \non \\ 
                         &\simeq&
    e^{-K} \frac{g^2}{\lambda^2} \frac{\overline{X}}{X} 
            \lmk X^2-\overline{X}^2 \rmk.   
\eea     
Here we have used the fact that $X\overline{X} - \mu'^2 = \CO(e^{-K})$
and $X \gg \overline{X}$. The F-term contribution to the potential
is then estimated as
\bea
  V_F &=& \frac{\overline{X} 
                \lmk X\overline{X} - \mu'^2 \rmk}
         {2X + \overline{X} \lmk X\overline{X} - \mu'^2 \rmk}
        g^2 \lmk X^2-\overline{X}^2 \rmk  \non \\
      &<& g^2 \lmk X^2-\overline{X}^2 \rmk
      \ll V_D = \frac{g^2}{8} \lmk X^2-\overline{X}^2 \rmk^2
\eea
for $X \gg 1$ and $\overline{X} \ll 1$. The potential thus is dominated
by the D-term during inflation. 

We also consider the mass matrix of the
fields $X$ and $\overline{X}$, $V_{,ij}|_{T}$($V_{,ij} \equiv \del^2
V_{\rm eff}/(\del\phi_i\del\phi_j)$), whose elements are given by
\bea
  V_{,XX}|_{T} &\simeq& 
      \frac{\lambda^2}{2} e^K \overline{X}^2
      + g^2 \lmk X^2-\overline{X}^2 \rmk \overline{X}^2
      + \frac{g^2}{2} \lmk 3X^2-\overline{X}^2 \rmk, \non \\             
  V_{,X\overline{X}}|_{T} &\simeq& 
      \frac{\lambda^2}{2} e^K X\overline{X}
      + \frac{g^2}{2} \lmk X^2-\overline{X}^2 \rmk 
        \frac{\overline{X}}{X} \lmk 1 + X^2 + \overline{X}^2 \rmk
      - g^2 X \overline{X}, \non \\              
  V_{,\overline{X}\,\overline{X}}|_{T} &\simeq& 
      \frac{\lambda^2}{2} e^K X^2
      + g^2 \lmk X^2-\overline{X}^2 \rmk \overline{X}^2
      + \frac{g^2}{2} \lmk 3\overline{X}^2-X^2 \rmk,
  \label{eq:matrix12}
\eea
up to the order of $\CO((e^{K})^0)$.\footnote{More precisely, we have
used the approximation that $|X\overline{X} - \mu'^2| \ll
X/\overline{X},\, X\overline{X}$.} The effective squared masses, say
$m^2$, of the fields $X$ and $\overline{X}$ are given as the solutions of the
following equation
\beq
  m^4 - (V_{,XX} + V_{,\overline{X}\,\overline{X}}) m^2 
    + V_{,XX} V_{,\overline{X}\,\overline{X}} - V_{,X\overline{X}}^2 = 0,
\eeq
where
\bea
  V_{,XX} + V_{,\overline{X}\,\overline{X}}|_{T} 
    &\simeq& \frac12 \lambda^2 e^{K} \lmk X^2 + \overline{X}^2 \rmk 
     \simeq  \frac12 \lambda^2 e^{K} X^2,
                      \non \\
  V_{,XX} V_{,\overline{X}\,\overline{X}} - V_{,X\overline{X}}^2|_{T} &\simeq& 
    \frac34 g^2 \lambda^2 e^{K} X^4,                        
  \label{eq:lambda}
\eea     
up to the order of $\CO(e^{K})$. Here, we have used the approximation
that $X\overline{X} - \mu'^2 = \CO(e^{-K})$ and $X \gg 1 \gg
\overline{X}$. The effective squared masses are then approximately given
by
\beq
  m^2 \simeq \frac12 \lambda^2 e^{K} X^2, \qquad \frac32 g^2 X^2 \ll H^2
  \simeq V_D / 3,
\eeq
where $H$ is a Hubble parameter. The inflaton field in the new D-term
chaotic inflation under discussion corresponds to this effectively
massless mode. This light squared mass vanishes for $g=0$ as expected,
reflecting the exact F-flat direction.

Since $V_{,\overline{X}\,\overline{X}} \gg V_{,XX}$ and $X \gg
\overline{X}$, the inflationary trajectory is given by the minimum of
the field $\overline{X}$ represented by $\del V / \del \overline{X} = 0$, 
and we can
 write the minimum of $\overline{X}$ as a function of $X$,
$\overline{X}^m = \overline{X}^m(X)$.  The field trajectory governing the
inflation dynamics therefore can be parameterized by the field $X$
which we call an inflaton.\footnote{Note here that the effectively
massless field trajectory parameterized by the inflaton field $X$
is different from the $X$ direction with the mass $V_{,XX} \gg
H^2$.} Then, by inserting the above relation into the effective
potential, we define the reduced  potential $V_{\rm r}(X)$ as
\beq
  V_{\rm r}(X) \equiv V_{\rm eff}(X,\overline{X}^m(X))
                     \lmk \simeq \frac{g^2}{8} X^4 \rmk.  
\eeq
As explicitly shown in Ref. \cite{YY}, when there is only one massless
mode and the other modes are massive, the generation of adiabatic
density fluctuations as wells as the dynamics of the homogeneous mode is
completely determined by the reduced potential $V_{\rm r}(X)$. Indeed,
the equation of motion for the homogeneous mode of the inflaton $X$
along the rolling direction (T) is approximated as
\beq
  \left.
    \ddot{X} + 3 H \dot{X} + \frac{\del V_{\rm eff}}{\del X}
  \right|_{T}
  =
    \ddot{X} + 3 H \dot{X} 
      + \frac{dV_{\rm r}}{dX} 
  = 0,
 \label{eq:homo}
\eeq
where the dot represents time derivative. Thus, the dynamics of the
inflation with the inflaton $X$ can be estimated by using the reduced
potential $V_{\rm r}(X)$ as long as the inflation dynamics follows the
minimum of $\overline{X}$.

Next, we evaluate the primordial density fluctuations in the
longitudinal gauge. The equations of motion for the perturbation $\delta
X$ and $\delta \overline{X}$ are given by \cite{fluc}
%
\bea
  && \ddot{\delta X} + 3 H \dot{\delta X} 
       - \frac{\nabla^2}{a^2} \delta X 
       + V_{,XX}|_{T}\,\delta X 
       + V_{,X\overline{X}}|_{T}\,\delta\overline{X}
      = -2 \left. \frac{\del V_{\rm eff}}{\del X} 
                    \right|_{T} X + 4 \dot{X}\dot{\Phi}, \non \\
  && \ddot{\delta \overline{X}} + 3 H \dot{\delta \overline{X}} 
       - \frac{\nabla^2}{a^2} \delta \overline{X} 
       + V_{,\overline{X}X}|_{T}\,\delta X 
       + V_{,\overline{X}\,\overline{X}}|_{T}\,\delta\overline{X}
      = -2 \left. \frac{\del V_{\rm eff}}{\del \overline{X}} 
         \right|_{T} \overline{X} + 4 \dot{\overline{X}}\dot{\Phi}, \non \\
  \label{eq:multifluc}
\eea
%
where $\Phi$ is the gravitational potential. We hereafter use the same
symbols $X$ and $\overline{X}$ for both the homogeneous modes and the
full fields for the notational brevity unless stated otherwise.

We are interested only in the adiabatic density fluctuations
characterized by the condition
\beq
  \frac{\delta X}{\dot{X}} = \frac{\delta\overline{X}}{\dot{\overline{X}}}
  \quad \Longleftrightarrow \quad
  \delta\overline{X} = \frac{d\overline{X}^m(X)}{dX} \delta X
  \label{eq:ad}
\eeq
where we have used
$\dot{\overline{X}}/\dot{X}=d\overline{X}^m(X)/dX$. Since the relation
$\del V_{\rm eff}(X,\overline{X}^m(X)) / \del \overline{X} = 0$ holds for
any $X$ in the regime of our interest, we find
\beq
  \left.
  \frac{d}{dX} \lkk 
      \frac{\del V_{\rm eff}}{\del \overline{X}}(X,\overline{X}^m(X))
                    \rkk
  = V_{,X\overline{X}} 
     + V_{,\overline{X}\,\overline{X}} \frac{d\overline{X}^m}{dX} \right|_{T}
  = 0.
  \label{eq:relation} 
\eeq

Then, using this relation, the equation of motion for the
perturbation $\delta X$ can be rewritten as
\beq
  \ddot{\delta X} + 3 H \dot{\delta X} 
       - \frac{\nabla^2}{a^2} \delta X + 
        \left.
          \frac{V_{,XX}V_{,\overline{X}\,\overline{X}}
         -V_{,X\overline{X}}^2}{V_{,\overline{X}\,\overline{X}}} 
                    \right|_{T} \delta X
      \left.
      = -2 \frac{\del V_{\rm eff}}{\del X} 
                    \right|_{T} \Phi + 4 X \dot{\Phi}. 
  \label{eq:fluc1}
\eeq
Taking into account this relation and
\beq
  \frac{d^2V_{\rm r}}{dX^2}
   = \left. V_{,XX} + 2 \frac{d\overline{X}^m}{dX} V_{,X\overline{X}}
     + \lmk \frac{d\overline{X}^m}{dX} \rmk^2 V_{,\overline{X}\,\overline{X}} 
        \right|_{T} 
   = \left. \frac{V_{,XX}V_{,\overline{X}\,\overline{X}}
              -V_{,X\overline{X}}^2}{V_{,\overline{X}\,\overline{X}}} \right|_{T}
   \lmk \simeq \frac32 g^2 X^2 \rmk,
\eeq
the equation of motion for the perturbation $\delta X$ becomes
\bea
  \ddot{\delta X} + 3 H \dot{\delta X} 
       - \frac{\nabla^2}{a^2} \delta X + 
        \frac{d^2V_{\rm r}}{dX^2} \delta X
      = -2 \frac{dV_{\rm r}}{dX} \Phi + 4 X \dot{\Phi}. 
  \label{eq:fluc2}
\eea
The perturbation $\delta X$ is hence completely determined by the
reduced potential $V_{\rm r}(X)$. Note that $d^2V_{\rm r}/dX^2$ is the
effective mass squared along the rolling direction given by $\del V_{\rm
eff} / \del \overline{X} = 0$, and this rolling direction
actually coincides with the eigenvector of the effectively massless mode
of $\lambda$.

On the other hand, using the adiabatic condition, the gravitational
potential is described only by $\delta\phi_1$ as
\bea
  \lmk \dot{H} - \frac{\nabla^2}{a^2} \rmk \Phi =
    \frac{1}{2M_G^2} 
       \lmk \ddot{X} \delta X - \dot{X} \dot{\delta X} \rmk 
       \lkk 1 + \lmk \frac{d\overline{X}^m}{dX} \rmk^2 \rkk.
 \label{eq:gpdirect} 
\eea
Consequently, the relation
\beq
 \dot{H} = - \frac{\dot{X}^2+\dot{\overline{X}}^2}{2M_G^2} 
         = - \frac{\dot{X}^2}{2M_G^2} 
             \lkk 1 + \lmk \frac{d\overline{X}^m}{dX} \rmk^2 \rkk,
\eeq
leads to the gravitational potential in the long wave limit
\beq
  \Phi = \frac{d~}{dt} \lmk \frac{\delta X}{\dot{X}}\rmk.
  \label{eq:gravitationalpot}
\eeq
This expression of the gravitational potential coincides with that of
the single field inflation with the reduced potential $V_{\rm r}(X)$. We
can as a result calculate the density fluctuations of our inflation
model based on the reduced potential $V_{\rm r}(X) \simeq g^2 X^4 / 8$,
which implies that the gauge coupling $g$ should be $g \sim 10^{-6}$ in
order to explain the primordial density fluctuations.

After inflation, the inflaton oscillates around the global minimum $X = \overline{X}
= \mu'$ and $S = 0$. Note that $S$ remains to vanish. 
Though the inflaton rolls down almost along the direction of
$X$ during inflation, the trajectory after inflation is curved so that
both $X$ and $\overline{X}$ oscillate around the global minimum. 
The mass matrix around the minimum is given by
\bea
  V_{,XX}|_{M} &=& 
    \frac12 \lambda^2 e^{K} \mu'^2 + g^2 \mu'^2, \non \\
  V_{,X\overline{X}}|_{M} &=& 
    \frac12 \lambda^2 e^{K} \mu'^2 - g^2 \mu'^2, \non \\
  V_{,\overline{X}\,\overline{X}}|_{M} &=& 
    \frac12 \lambda^2 e^{K} \mu'^2 + g^2 \mu'^2,
  \label{eq:matrixmin}
\eea
where the suffix $M$ represents the evaluation at the global minimum
$M$. Hence, the effective squared masses $m_{\pm}^2$ are given by
$m_{+}^2 = \lambda^2 e^K \mu'^2$ with the eigendirection $X_{+} \equiv
(X+\overline{X})/\sqrt{2}$ and $m_{-}^2 = 2 g^2 \mu'^2$ with the
eigendirection $X_{-} \equiv (X-\overline{X})/\sqrt{2}$.

\section{Reheating and Leptogenesis in New D-term Chaotic Inflation}

\label{sec:lepto}

Let us now discuss the issues of the reheating and 
the baryon asymmetry of the universe in the new D-term chaotic inflation 
through a concrete example, namely, the non-thermal leptogenesis scenario via
the inflaton decay which produces the the heavy Majorana neutrinos $N_{i}$ 
non-thermally \cite{LGinfdec}.

We consider the right handed Majorana neutrinos
in addition to 
the Minimal Supersymmetric Standard Model (MSSM) fields in the 
superpotential
\beq
 W = \lambda S (X \overline{X} - \mu^2) 
    + \sum_{i} \alpha_i X \overline{X} N_i N_i
    + \sum_{i,j} h^{\nu}_{ij} N_i L_j H_u
    + W_{\rm MSSM},
 \label{eq:Nsuperpotential}
\eeq
where the subscripts $i$ and $j$ represent the generation indices,
$h^{\nu}_{ij}$ is the Yukawa coupling, $L_j$ is the lepton doublet,
$H_u$ is the up-type Higgs and $W_{\rm MSSM}$ is the superpotential of
the MSSM. The charges for various supermultiplets are given in Table
\ref{tab:charges}. Note that non-renormalizable terms like
$N_iN_i(X\overline{X})^n$ and $S(X\overline{X})^n$ can appear in the
superpotential but they are negligible as long as $\mu \ll 1$ because
$|X\overline{X}| \sim \mu^2$ during inflation.

Taking the canonical K\"ahler potential, the minimal gauge kinetic
function and the vanishing FI term, we can calculate the scalar
potential consisting of the F-term $V_F$ and D-term $V_D$
\bea
  V &=& V_F + V_D, \non \\ 
  V_F  &=& e^{K} \lkk \,
          \biggl| \lambda \lmk X\overline{X} 
                 - \mu^2 \rmk + S^{\ast} W  \biggr|^2
        + \biggl| \lambda S \overline{X} 
                 + \sum{i} \alpha_i \overline{X} N_i N_i 
                 + X^{\ast} W  \biggr|^2  \right. \non \\
       && + \biggl| \lambda S X 
                 + \sum{i} \alpha_i X N_i N_i 
                 + \overline{X}^{\ast} W  \biggr|^2
        + \sum_i \biggl| 2 \alpha_i X \overline{X} N_i
                 + \sum_{j} h^{\nu}_{ij} L_j H_u 
                 + N_{i}^{\ast} W  \biggr|^2 \non \\ 
       && \left.
        + \sum_k \biggl| \frac{\del W}{\del \xi_k}
                 + \xi_{k}^{\ast} W  \biggr|^2 
        - 3 |W|^2 \rkk   
        , \non \\
  V_D&=& \frac{g^2}{2} 
       \lmk |X|^2 - |\overline{X}|^2 \rmk^2 + V_D^{\rm MSSM},
\eea
where $\xi_k$ represents the MSSM field and $V_D^{\rm MSSM}$ represents
the D-term concerning the standard model gauge group. The minimum of the
F-term (the F-flat condition) is given by
\beq
  V_F = 0 \,\, \Longleftrightarrow \,\, 
        X\overline{X} - \mu^2 = 0 \quad\&\quad S = 0
        \quad\&\quad N_{i} = 0 \quad\&\quad 
        {\rm MSSM\,\,F-flat\,\,condition}. 
\eeq
On the other hand, the minimum of the D-term (the D-flat condition) is
given by
\beq
    V_D = 0\,\, \Longleftrightarrow \,\,
    |X| = |\overline{X}| \quad\&\quad {\rm MSSM\,\,D-flat\,\,condition}.
\eeq
The global minimum of the potential hence is given by
\beq
  S = 0,\,\,\, X = \mu e^{i\theta},\,\,\, \overline{X} = \mu e^{-i\theta},
  \,\,\, {\rm MSSM\,\,flat\,\,condition}
\eeq
where the phase $\theta$ is arbitrary.
As was done in the previous section, one can show that $S$ and
$N_i$ go to the zeros dynamically during the inflation so that the effective
dynamics is described by the effective potential
\beq
  V_{\rm eff}(X,\overline{X}) = 
     \frac{\lambda^2}{4} e^{K} 
        \lmk X\overline{X} - \mu'^2 \rmk^2 
      + \frac{g^2}{8} \lmk X^2 - \overline{X}^2 \rmk^2,
\eeq
where we have used the redefined fields $X \equiv \sqrt{2}\, {\rm Re}\,X$
and $\overline{X} \equiv \sqrt{2}\, {\rm Re}\,\overline{X}$. Thus, the
dynamics and primordial fluctuations of the inflation are essentially
unchanged even if we include other fields besides X and $\overline{X}$.

Let us now discuss the reheating and leptogenesis in this model. Note
that the spontaneous breaking of the gauge symmetry due to the non-vanishing
VEV of the inflaton is crucial because, if the inflaton VEV vanishes, the
charge conservation would prohibit the inflaton decay in our model.
The Majorana masses of right-handed neutrinos $M_{i}$ are given by 
$M_{i} = \alpha_{i}\la X\overline{X} \ra / 2 = \alpha_{i} \mu'^2/2$. 
The inflatons $X$ and 
$\overline{X}$ (or equivalently $X_{+}$ and $X_{-}$) can decay into 
the right handed neutrinos $N_{i}$ through the Yukawa
interactions if $M_i < m_{+}/2\,\,{\rm or}\,\, m_{-}/2$. 
Let us now, for concreteness, consider the case where
 $\lambda \gg g$ and the decay of
$X_{-}$ to the right handed neutrinos is kinematically prohibited. 
The partial decay width of the inflaton to the right handed neutrinos 
$X_+\rightarrow N_i N_i, \Gamma_{(X_{+} \rightarrow N_{i})}$, is then given by
\bea
  \Gamma_{(X_{+} \rightarrow N_{i})} 
     &\simeq& 
        \frac{1}{32\pi}
        \alpha_i^{2} \la X \ra^2 m_{+} 
     \sim 
        \frac{1}{32\pi} \alpha_i^{2} \lambda \mu'^3 \non \\
     &\sim& 10^{-3}~{\rm GeV} 
          \lmk \frac{\alpha_i}{0.1} \rmk^{2}
          \lmk \frac{\lambda}{10^{-4}} \rmk
          \lmk \frac{\mu'}{10^{14}~\rm GeV} \rmk^3.
  \label{eq:Xdecay}
\eea
The reheating temperature $T_R$ hence becomes
\beq
  T_R \simeq 0.1 \sqrt{  \Gamma_{(X_{+} \rightarrow N_{i})} }
      \sim 10^{7}~{\rm GeV} 
           \lmk \frac{\alpha}{0.1} \rmk
           \lmk \frac{\lambda}{10^{-4}} \rmk^{\frac12}
           \lmk \frac{\mu'}{10^{14}~\rm GeV} \rmk^{\frac32}
\eeq
with $\alpha \equiv \sqrt{\sum_i{\alpha_i^2}}$ where the sum $i$ runs
for only the generations of right-handed neutrinos which the inflaton
$X_{+}$ can decay into. 

The produced $N_{i}$ decay into leptons $L_{j}$ and Higgs doublets
$H_{u}$ through the Yukawa interactions via
\beq
  W = h_{\nu}^{ij}N_{i}L_{j}H_{u}
\eeq
where we have taken a basis where the mass matrix for $N_{i}$ is 
diagonal. We also assume $|(h_{\nu})_{i3}| > |(h_{\nu})_{i2}| \gg
|(h_{\nu})_{i1}|$ (i = 1, 2, 3). We consider only the decay of $N_{1}$
assuming that the mass $M_{1}$ is much smaller than the others, $M_{2}$
and $M_{3}$.  
The interference between the tree-level and the one-loop
diagrams including vertex and self-energy corrections generates the lepton
asymmetry \cite{FY,CRV,FPS,BP},

\bea
   \label{eqn:cpasym}
  \epsilon_1 
    &\equiv&
    \frac{ \Gamma (N_1 \rightarrow H_u + l )
         - \Gamma (N_1 \rightarrow \overline{H_u} + \overline{l} ) }
         { 
\Gamma (N_1 \rightarrow H_u + l )
         +\Gamma (N_1 \rightarrow \overline{H_u} + \overline{l} ) 
}
    \non \\
    &=& - 
    \frac{3}{ 8 \pi \left( h_\nu h_\nu^{\dagger} \right)_{11} }
    \sum_{i=2,3} 
        \mbox{Im} \left( h_\nu h_\nu^{\dagger} \right)_{1i}^2 
        \frac{M_1}{M_i}
\non
\\
&\simeq& \frac{3}{8\pi} \frac{M_1}{\langle H_u\rangle^2}m_{\nu_3}
\delta_{\rm eff}
\non 
\\
&\sim&10^{-7} \left(\frac{M_1}{10^9 \rm GeV}\right)
\left(\frac{m_{\nu_3}}{10^{-2} \rm eV}\right) \delta_{\rm eff},
\eea
where the effective CP violation phase is
\bea
  \delta_{\rm eff}\equiv -\frac{ \mbox{Im} 
     \left[ h_\nu (m^*_{\nu}) h_\nu^{T} \right]_{11}^2 }
    {m_{\nu_3}\left( h_\nu h_\nu^{\dagger} \right)_{11}}.
\eea
$m_{\nu_{3}}$ here is a mass eigenvalue of the left-handed neutrino mass 
matrix $m_{\nu}$ estimated by the seesaw mechanism
\cite{seesaw} as, based on our assumption that the $(h_{\nu})_{33}$ is the dominant
entry in $h_{\nu}$ and $M_3\gg M_1$,
\bea
  m_{\nu_{3}} &\simeq& \frac{\left|\lmk h_\nu \rmk_{33}\right|^2
                           \la H_{u} \ra^{2}}{M_{3}} \non \\
              &\sim& 10^{-2}~{\rm eV} \lmk 
                       \frac{\left|\lmk h_\nu\rmk_{33}\right|}
                            {5 \times 10^{-3}} \rmk^2
                        \lmk \frac{M_{3}}{10^{10}~\rm GeV} \rmk^{-1},
\eea
which is consistent with the mass inferred from the Super-Kamiokande
experiments \cite{SK} for $|\lmk h_\nu\rmk_{33}| \sim 10^{-2}$
and $M_{3} \sim 10^{10}$~GeV.

The total decay rate of $N_{1}$, $\Gamma_{N_1}$, is given by
\bea
  \Gamma_{N_1} &=& \Gamma (N_1 \rightarrow H_u + l )
         + \Gamma (N_1 \rightarrow \overline{H_u} + \overline{l} )
                   \non \\
               &\simeq& \frac{1}{8\pi} \Sigma |(h_{\nu})_{1i}|^{2}
                         M_{1} \non \\
               &\simeq& \frac{1}{8\pi} |(h_{\nu})_{13}|^{2} M_{1} 
                         \non \\
               &\sim& 10~{\rm GeV} 
                      \lmk \frac{|(h_{\nu})_{13}|}
                                {5 \times 10^{-4}} \rmk^{2}
                      \lmk \frac{M_{1}}{10^{9}~\rm GeV} \rmk.
\eea
Thus, for a wide range of parameters, the decay rate $\Gamma_{N_1}$ is
much larger than the decay rate of the inflaton $\Gamma_{(X_{+}
\rightarrow N_{i})}$ so that the produced $N_{1}$ immediately decays
into leptons and Higgs supermultiplets.

Before estimating the lepton asymmetry produced in our model, let us
evaluate the lepton asymmetry needed to explain the observed baryon
number density. A part of the produced lepton asymmetry (or, exactly speaking, $B-L$
asymmetry) is converted into the baryon asymmetry through the
sphaleron processes, which can be estimated as \cite{KSHT}
\beq
  \frac{n_B}{s} \simeq - \frac{8}{23} \frac{n_L}{s},
\eeq
where we have assumed the standard model with two Higgs doublets and
three generations. In order to explain the observed baryon number
density,
\beq
  \frac{n_B}{s} \simeq (0.1 - 1) \times 10^{-10},
\eeq
we need the lepton asymmetry,
\beq
  \frac{n_L}{s} \simeq - (0.3 - 3) \times 10^{-10}.
\eeq

Now we estimate the lepton asymmetry produced through the inflaton
decay. For $M_{1} \gtrsim 10^{9}$~GeV, $M_{1}$ is one hundred
times larger than the reheating temperature $T_{R}$. In this
case, the produced $N_{1}$ is out of equilibrium and the
ratio of the lepton number to entropy density can be estimated as
\bea
  \frac{n_L}{s} 
          &\simeq&
       \frac32~\epsilon_1 B_r \frac{T_R}{m_{+}} 
    \non \\
    &\sim& 
    - 10^{-10} \delta_{\rm eff} B_r
    \lmk \frac{M_1}{10^{9}~\mbox{GeV}} \rmk
   \lmk \frac{T_R}{10^{7}~\mbox{GeV}} \rmk
    \lmk \frac{m_{+}}{10^{10}~\mbox{GeV}} \rmk^{-1}
    \non \\
    &\sim& 
    - 10^{-10} \delta_{\rm eff} B_r
    \lmk \frac{\lambda}{10^{-4}} \rmk^{-\frac12}
    \lmk \frac{\alpha}{0.1} \rmk
    \lmk \frac{\alpha_1}{0.1} \rmk
    \lmk \frac{\mu'}{10^{14}~\mbox{GeV}} \rmk^{\frac52},
\eea
where $B_r$ is the branching ratio of the inflaton $X_{+}$ into $N_1$.
For $M_{3} \sim M_{2} \sim m_{+} \sim 10^{10}$~GeV with $\alpha_2 \sim
\alpha_3 = \CO(1)$, $m_{+}$ is comparable to $M_2$ and $M_3$ so that the
decay into $N_{2}$ and $N_{3}$ are prohibited kinematically or
suppressed by the phase space and hence $B_{r} \approx \CO(1)$. 
Note also that since $m_{-} \sim 10^8$~GeV for $g \sim 10^{-6}$ and
$\mu' \sim 10^{14}$~GeV, $m_{-} \ll M_1, M_2, M_3$ so that the decays of
$X_{-}$ into all the right-handed neutrinos are prohibited
kinematically, as assumed previously. In this case, $\alpha = \alpha_1
\sim 0.1$ and $T_R \sim 10^7$~GeV (low enough reheating temperature to avoid the
gravitino problem \cite{kawa}) resulting in $n_L/s \sim 
10^{-10} \delta_{\rm eff}$ which is consistent with the baryon number
density in the present universe.

\section{Summary and discussion}

\label{sec:con}

In this paper we have proposed a new model of D-term dominated chaotic
inflation in supergravity. The F-flat direction present in this model is
automatically lifted by the D-term, which leads to chaotic
inflation. The superpotential of our model was originally proposed as
the F-term hybrid inflation where the gauge singlet field $S$ plays the
role of an inflaton with the vanishing $X$ and $\overline{X}$ during
inflation.  On the other hand, we showed that another initial condition
such that $S \sim 0$, $|X| \gg 1$ or $|\overline{X}| \gg 1$ with
$X\overline{X} \sim \mu^2$ can naturally occur around the Planck scale
for a successful D-term chaotic inflation.  In contrast to the
previously proposed D-term chaotic inflation model \cite{KY}, the
inflaton can acquire the non-vanishing VEV after inflation which breaks
the gauge symmetry spontaneously so that it can decay into the visible
sector for the sufficient reheating\footnote{It was recently pointed out
that the supergravity effects induce the inflaton decay into the visible
sector when the inflaton acquires the non-vanishing VEV after inflation
\cite{idecay}, and an analogous mechanisms could help reheat the
universe for a D-term chaotic inflation model as well.}.  No cosmic
string is formed after inflation because the U(1) gauge symmetry is
broken during inflation, while such a cosmic string formation in the
conventional D-term inflation is often problematic \cite{prob}.

Our model leads to the quartic potential chaotic inflation which has the
tight constraints from the recent observations \cite{WMAP3}. The
relaxation of such constraints is possible by, for instance, an
appropriate choice of the non-minimal gauge kinetic function such as a
form $f = 1 + d_X |X|^2 + d_{\overline{X}} |\overline{X}|^2$
($d_X,d_{\overline{X}}$ : constants) which gives a quadratic inflaton
potential. One of the present authors (T.K.)  also proposed the
quadratic potential chaotic inflation by use of the FI field
\cite{kawano} even though a successful reheating needs more care
\cite{reheating}. The consideration of the primordial fluctuations from
a MSSM flat direction acting as a curvaton in our model could be another
possibility for a viable quartic potential chaotic inflation model
\cite{MTT,HKMT}. We also mention that, for a toy model using a minimal gauge kinetic function illustrated in our discussion, the gauge coupling $g$ should
be $g \sim 10^{-6}$ in order to explain the primordial density
fluctuations. This value of the gauge coupling is much smaller than the
standard gauge couplings. However, this may not be a problem because the
gauge symmetry may be a hidden gauge symmetry, or the gauge coupling
could be suppressed, for instance, by considering the extra dimensions.
This topic would be worth further investigation.

We have also discussed the leptogenesis scenario via the inflaton decay
in this chaotic inflation model, which can explain the observed baryon
asymmetry for a reasonable parameter set consistent with the data from the 
neutrino experiments.

\subsection*{Acknowledgments}

We thank A. D. Linde, R. Kallosh, Y. Shinbara, F. Takahashi, and
J. Yokoyama for useful discussions.  K.K. is supported by DOE grant
DE-FG02-94ER-40823.  T.K. was supported in part by a Grant-in-Aid
(\#19540268) from the MEXT of Japan.  M.Y. is supported in part by JSPS
Grant-in-Aid for Scientific Research No.~18740157 and No.~19340054.


\begin{thebibliography}{99}

\bib{chaotic} 
A. D. Linde, 
\PLBold{129}{177}{83}.

\bib{inflation}
A.D. Linde, {\it Particle Physics and Inflationary Cosmology}
(Harwood, Chur, Switzerland, 1990);
A. R. Liddle and D. H. Lyth, 
{\it Cosmological Inflation and Large Scale Structure}  
(Cambridge University Press, Cambridge, England 2000); 
D. H. Lyth and A. Riotto,
\PRT{314}{1}{99}.

\bib{hi}
G. Lazarides, C. Panagiotakopoulos, and N. D. Vlachos,
\PRD{54}{1369}{96};
G. Lazarides and N. D. Vlachos,
\IBID{56}{4562}{97};
N. Tetradis,
\IBID{57}{5997}{98}.

\bib{SUSY}
See, for a review, H. P. Nilles,
\PRT{110}{1}{84}.

\bib{GL}
A. S. Goncharov and A. D. Linde,
\PLBold{139}{27}{84}; \CQG{1}{L75}{84}.

\bib{MSYY}
H. Murayama, H, Suzuki, T. Yanagida, and J. Yokoyama,
\PRD{50}{R2356}{94}.

\bib{KYY}
M. Kawasaki, M. Yamaguchi, and T. Yanagida,
\PRLL{85}{3572}{00}.

\bib{KYY2}
M. Kawasaki, M. Yamaguchi, and T. Yanagida,
\PRDD{63}{103514}{01}.

\bib{variSG}
M. Yamaguchi and J. Yokoyama,
\PRDD{63}{043506}{01};
\IBB{68}{123520}{03}; 
M. Yamaguchi,
\IBIDD{64}{063502}{01};
\IBB{64}{063503}{01};
M. Kawasaki and M. Yamaguchi,
\IBIDD{65}{103518}{02}.

\bib{variSG2}
S. C. Davis and M. Postma,
\JCAPP{03}{015}{08}.

\bib{variSS}
J. P. Hsu, R. Kallosh, and S. Prokushkin,
\JCAPP{12}{009}{03};
H. Firouzjahi and S. H. H. Tye,
\PLBB{584}{147}{04};
J. P. Hsu, R. Kallosh,
\JHEPP{04}{042}{04};
R. Kallosh, N. Sivanandam, and M. Soroush,
\PRDD{77}{043501}{08}. 

\bib{dtermorig}
E. D. Stewart,
\PRD{51}{6847}{95};
P. Binetruy and G. Dvali,
\PLB{388}{241}{96};
E. Halyo,
\PLB{387}{43}{96}.

\bib{dterm}
J. Rocher and M. Sakellariadou,
\PRLL{94}{011303}{05};
\JCAPP{03}{004}{05};
\IBB{11}{001}{06};
O. Seto and J. Yokoyama,
\PRDD{73}{023508}{06}.

\bib{KY}
K. Kadota and M. Yamaguchi,
\PRDD{76}{103522}{07}.

\bib{FY}M. Fukugita and T. Yanagida,
\PLB{174}{45}{86}.

\bib{LGinfdec}
K. Kumekawa, T. Moroi, and T. Yanagida,
\PTP{92}{437}{94}; 
G. Lazarides, Springer Tracts Mod. Phys. {\bf 163} 227 (2000), and reference therein.
G. F. Giudice, M. Peloso, A. Riotto, and I. Tkachev,
\JHEP{08}{014}{99};
T. Asaka, K. Hamaguchi, M. Kawasaki, and T. Yanagida,
\PLB{464}{12}{99}; \PRDD{61}{083512}{00}.

\bib{KK}
R. Kallosh, L. Kofman, A. D. Linde, and V. A. Proeyen,
\CQGG{17}{4269}{00};
P. Binetruy, G. Dvali, R. Kallosh, and V. A. Proeyen,
\CQGG{21}{3137}{04};
H.~Elvang, D.~Z.~Freedman and B.~Kors,
JHEP {\bf 0611}, 068 (2006).

\bib{try}
G.~Villadoro and F.~Zwirner,
Phys.\ Rev.\ Lett.\  {\bf 95}, 231602 (2005);
C.~P.~Burgess, R.~Kallosh and F.~Quevedo,
JHEP {\bf 0310}, 056 (2003).

\bib{Fhybrid}
E. J. Copeland, A. R. Liddle, D. H. Lyth, E.D. Stewart, and D. Wands,
\PRD{49}{6410}{94};
G. Dvali, Q. Shafi, and R. K. Shaefer,
\PRL{73}{1886}{94};
G. Lazarides, R. K. Schaefer, and Q. Shafi,
\PRD{56}{1324}{97};
R. Jeannerot,
\PRD{53}{5426}{96};
A. Linde and A. Riotto, 
\PRD{56}{R1841}{97}.

\bib{hybinit}
L .E. Mendes and A. R. Liddle,
\PRDD{62}{103511}{00};
N. Tetradis,
\IBID{57}{5997}{98}.

\bib{YY}
M. Yamaguchi and J. Yokoyama,
\PRDD{74}{043523}{06}.

\bib{fluc}
J. M. Bardeen,
\PRD{22}{1882}{80};
H. Kodama and M. Sasaki,
\PTPS{78}{1}{84};
V.F. Mukhanov, H.A. Feldman, and R. H. Brandenberger,
\PRT{215}{203}{92}.

\bib{CRV}
L. Covi, E. Roulet, and F. Vissani,
\PLB{384}{169}{96}.

\bib{FPS}
M. Flanz, E. A. Paschos, and U. Sarkar,
\PLB{345}{248}{95};
\IB{384}{487(E)}{96}.

\bib{BP}
W. Buchm\"uller and  M. Pl\"umacher,
\PLB{431}{354}{98}.

\bib{seesaw}
T. Yanagida, 
in Proceedings of the Workshop on the Unified Theory and 
the Baryon Number of the Universe, Tsukuba, Japan, 1979, 
edited by O. Sawada and S. Sugamoto,
(KEK, Tsukuba, 1979);
M. Gell-Mann, P. Ramond, and R. Slansky,
in {\it Supergravity}
edited by D.Z. Freedman and P. van Nieuwenhuizen 
(North-Holland, Amsterdam, 1979).

\bib{SK}
Super-kamiokande Collaboration, Y. Fukuda {\it et al.},
\PLB{433}{9}{98}; \IB{436}{33}{98}; \PRL{81}{1562}{98}.

\bib{KSHT}
S. Y. Khlebnikov and M. E. Shaposhnikov,
\NPB{308}{885}{88}; \\
J. A. Harvey and M. S. Turner,
\PRD{42}{3344}{90}.

\bibitem{kawa}
  M.~Kawasaki, K.~Kohri, and T.~Moroi,
  Phys.\ Rev.\  D {\bf 71}, 083502 (2005).
 
  M.~Kawasaki, K.~Kohri, and T.~Moroi,
  Phys.\ Lett.\  B {\bf 625}, 7 (2005).

\bib{idecay}
M. Endo, M. Kawasaki, F. Takahashi, and T. T. Yanagida,
\PLBB{642}{518}{06};
M. Endo, K. Kadota, K. A. Olive, F. Takahashi, and T. T. Yanagida,
\JCAPP{02}{018}{07}; 
M. Endo, F. Takahashi, and T. T. Yanagida,
\PRDD{76}{083509}{07};
  T.~Asaka, S.~Nakamura and M.~Yamaguchi,
  Phys.\ Rev.\  D {\bf 74}, 023520 (2006).

\bibitem{prob}
  D.~H.~Lyth and A.~Riotto,
  Phys.\ Lett.\  B {\bf 412}, 28 (1997);

  J.~Rocher and M.~Sakellariadou,
  \JCAPP{11}{001}{06}.


\bib{WMAP3}
D. N. Spergel {\it et al.},
\APJSS{170}{377}{07}.

\bib{kawano}
T. Kawano,
arXiv:0712.2351.

\bib{reheating}
T. Kawano, work in progress.

\bib{MTT}
T. Moroi, T. Takahashi, and Y. Toyoda,
\PRDD{72}{023502}{05};
T. Moroi, T. Takahashi,
\PRDD{72}{023505}{05}.

\bib{HKMT}
K. Hamaguchi, M. Kawasaki, T. Moroi, and F. Takahashi,
\PRDD{69}{063504}{04}.

\end{thebibliography}
\end{document}